\documentclass[prl,aps,showpacs,floatfix,nofootinbib,twocolumn,letterpaper,reprint]{revtex4-1}
\usepackage{epsfig}
\usepackage{amssymb}
\usepackage{amsmath}
\usepackage{amsfonts}
\usepackage{xcolor,graphicx}
\usepackage{bm}

\newcommand{\unit}{1\!\!1}
\newcommand{\Tr}{\mathrm{Tr}}

\newcommand{\ssec}[1]{\emph{#1}.---}

\begin{document}

\title{Pairing correlations across the superfluid phase transition in the unitary Fermi gas}
\author{S.~Jensen,$^1$ C. N.~Gilbreth,$^2$ and Y. Alhassid$^1$}
\affiliation{$^{1}$Center for Theoretical Physics, Sloane Physics Laboratory, Yale University, New Haven, CT 06520\\
$^{2}$Institute for Nuclear Theory,
Box 351550, University of Washington, Seattle, WA 98195} 

\begin{abstract}
In the two-component Fermi gas with a contact interaction, a pseudogap regime can exist at temperatures between the superfluid critical temperature $T_c$ and a temperature $T^* > T_c$.  This regime is characterized by pairing correlations without superfluidity. However, in the unitary limit of infinite scattering length, the existence of this regime is still debated.  To help address this, we have applied finite-temperature auxiliary-field quantum Monte Carlo (AFMC) to study the thermodynamics of the superfluid phase transition and signatures of the pseudogap in the spin-balanced homogeneous unitary Fermi gas.
We present results at finite filling factor $\nu \simeq 0.06$ for the condensate fraction, an energy-staggering pairing gap, the spin susceptibility,  and the heat capacity, and compare them to experimental data when available. In contrast to previous AFMC simulations, our model space consists of the complete first Brillouin zone of the lattice, and our calculations are performed in the canonical ensemble of fixed particle number.  The canonical ensemble AFMC framework enables the calculation of a model-independent gap, providing direct information on pairing correlations without the need for numerical analytic continuation.  We use finite-size scaling to estimate $T_c$ at the corresponding filling factor.  We find that the energy-staggering pairing gap vanishes above $T_c$, showing no pseudogap effects, and that the spin susceptibility shows a substantially reduced signature of a spin gap compared to previously reported AFMC simulations.
\end{abstract}

\maketitle

\ssec{Introduction}
The unitary Fermi gas (UFG) is the infinite-scattering-length limit of a system of spin-$1/2$ fermions with a zero-range interaction.  This system is relevant to a variety of physical systems, including neutron stars, strongly correlated QCD matter~\cite{Nishida2005} and high-$T_c$ superconductors~\cite{Randeria2010}.  The homogenous UFG is a strongly correlated quantum many-body system characterized by a single energy scale and is of broad interest as a testing ground for many-body theories.

The UFG has been realized experimentally using ultracold dilute gases of $^6$Li and $^{40}$K; see, e.g., Refs.~\cite{Ketterle2008, Zwierlein2017}. These experiments have measured various properties of the UFG, including the thermal energy, pressure, heat capacity, compressibility, and spectral function~\cite{Ku2012,Navon2009,Kinast2005,Schirotzek2008,Gaebler2010}. The UFG exhibits a superfluid phase transition at a critical temperature measured as $T_c = 0.167(13) T_F$~\cite{Ku2012} where $T_F$ is the Fermi temperature.

The nature of pairing correlations in the UFG above $T_c$ remains incompletely understood.  In particular, a pseudogap regime, in which pairing correlations exist even though a superfluid condensate is not present, was proposed to exist above $T_c$. Such a regime exists
in the BEC limit, where particles pair to form bound dimers at a temperature $T^*$ and condense at the critical temperature $T_c < T^*$.
However, in the UFG, it is still debated whether $T_c$ and the temperature scale $T^*$ for pairing should coincide or differ, and if they differ, what the properties of the pseudogap regime $T_c < T < T^*$ are.

A number of experimental works claimed to have observed signatures of a pseudogap in the UFG~\cite{Wulin2011,Gaebler2010}, while others have seen no signatures of a pseudogap~\cite{Ku2012,Navon2009,Navon2011,Sommer2011}.  Similar differences emerged in theoretical studies, with some showing a signature of a pseudogap~\cite{Magierski2009,Magierski2011,Wlazlowski2013,Chien2010a, Chien2010b,Perali2011,Kashimura2012,Palestini2012,Tajima2014,Wulin2011,Pini2019}, and others not~\cite{Haussmann2009, Enss2012}. For a recent review, see Ref.~\cite{Mueller2017}. A wide variety of theoretical methods were applied to study the superfluid phase transition of the UFG~\cite{Zwerger2012}.  While these methods provided important insight into the physics of the UFG, \textit{ab initio} simulations can provide the most accurate results~\cite{Burovski2006,Akkineni2007,Goulko2010,Drut2012,Van2012}. 

Here we apply finite-temperature auxiliary-field quantum Monte Carlo (AFMC) on a lattice to study the thermodynamic properties of the homogeneous UFG.  Our calculations differ from previous AFMC calculations~\cite{Magierski2009, Bulgac2008, Wlazlowski2013} in that (i) we do not use a spherical cutoff in the single-particle momentum space, but include the complete first Brillouin zone, leading to qualitatively different results, (ii) we use the canonical ensemble of fixed particle number, allowing the direct computation of a model-independent pairing gap from the staggering of energy in particle number without the need for numerically difficult analytic continuation,  (iii) we extrapolate to zero imaginary time step, and (iv) we calculate the heat capacity, which is challenging to compute in quantum Monte Carlo simulations. We also present results for the condensate fraction and static spin susceptibility.

Our calculations are done for a small but finite filling factor of $\nu \simeq 0.06$.  We use finite-size scaling of the condensate fraction to determine a critical temperature of  $T_c \simeq 0.130(15)T_{F}$ at this finite density.  We find that the model-independent pairing gap vanishes at temperatures larger than $T_c$, and thus does not show pseudogap effects, in contrast to the conclusions of Refs.~\cite{Magierski2009, Bulgac2008}.  The spin susceptibility shows a moderate suppression above $T_c$ and below a spin-gap temperature of $T^* \lesssim0.17\,T_F$, in contrast to the value of $T^* \simeq 0.25\,T_F$ found in Ref.~\cite{ Wlazlowski2013}.

Extensive supplemental material accompanies this article, in which we discuss important technical details~\cite{supp_material}.

\ssec{Lattice formulation and Hamiltonian} We consider $N$ spin-$1/2$ fermions that interact via a contact interaction $V=V_0 \delta ({\bf r}- {\bf r}')$ within a spatial volume with periodic boundary conditions. The volume is discretized into a lattice with an odd number $N_L$ of points in each dimension, each lattice point centered within a cube of side length $\delta x$.  The lattice Hamiltonian is
\begin{equation} \label{ham}
\hat{H}=\sum_{\mathbf{k},s}\epsilon _{\mathbf{k}}\hat{a}^{\dagger }_{\mathbf{k},s }\hat{a}_{\mathbf{k},s }+g\sum_{\mathbf{x}}\hat{n}_{\mathbf{x},\uparrow}\hat{n}_{\mathbf{x},\downarrow} \;,
\end{equation}
where $\hat{a}^{\dagger }_{\mathbf{k},s}$ and $\hat{a}_{\mathbf{k},s}$ are creation and annihilation operators for fermions with wavevector ${\bf k}$ and spin projection $s=\pm 1/2$, and the single-particle dispersion relation is $\epsilon _{\mathbf{k}} = {\hbar^2 \mathbf{k}}^{2} /2m$.  The coupling constant is $g=V_0/(\delta x)^{3}$ and $\hat{n}_{\mathbf{x},s}=\hat{\psi}^{\dagger}_{\mathbf{x},s}\hat{\psi}_{\mathbf{x},s}$, where $\hat{\psi}^{\dagger}_{\mathbf{x},s}, \hat{\psi}_{\mathbf{x},s}$ obey anticommutation relations $\{ \hat{\psi}^{\dagger}_{\mathbf{x},s},\hat{\psi}_{\mathbf{x}',s'}\}= \delta_{\mathbf{x},\mathbf{x}'}\delta_{s,s'}$.

Our single-particle model space consists of all single-particle states with spin projection $s=\pm 1/2$ and momentum ${\hbar\bf k}$ within the complete first Brillouin zone of the lattice, described by a cube $|k_i| \le k_c \,\,\, (i=x,y,z)$ with $k_c=\pi/\delta x$.  The thermodynamic limit of the UFG is recovered in the limits of zero filling factor ($\nu=N/N_L^3\rightarrow 0$) and large number of fermions ($N\rightarrow \infty$).

We choose $V_0$ to reproduce the scattering length $a$~\cite{Werner2012}
\begin{equation} \label{geqn}
\frac{1}{V_0}=\frac{m}{4\pi \hbar^2 a}-\int_{B}\frac{d^{3}k}{(2\pi)^{3}2\epsilon_{\mathbf{k}}} \;,
\end{equation}
which is derived by solving the Lippmann-Schwinger equation. We use the complete first Brillouin zone $B$ when calculating the integral in (\ref{geqn}). Solving the scattering problem numerically on the lattice, we find that Eq.~\eqref{geqn} is very accurate even for finite lattices: on the $9^3$ lattice it yields $a^{-1} = 0.006 \, (\delta x)^{-1}$ and an effective range of $r_e = 0.34 \, \delta x$~\cite{supp_material}, in close agreement with its value $r_e=0.337 \,\delta x$  in the limit of large lattices~\cite{Werner2012}.

\ssec{Finite-temperature AFMC}
The AFMC method (for a recent review, see Ref.~\cite{Alhassid2017}) is based on the Hubbard-Stratonovich (HS) transformation~\cite{Stratonovich1957,Hubbard1959}, which expresses the thermal propagator $e^{-\beta \hat H}$ ($\beta=1/k_BT$ is the inverse temperature $T$ with Boltzmann constant $k_B$) as a path integral over external auxiliary fields.

Dividing the imaginary time $\beta$ into $N_\tau$ imaginary times of length $\Delta \beta$, we use a symmetric Trotter decomposition 
\begin{equation}
e^{-\beta \hat{H}} = [e^{-\Delta \beta \hat H_0/2}e^{-\Delta \beta \hat{V}}e^{-\Delta \beta \hat H_0/2}]^{N_\tau} + O((\Delta \beta)^{2}) \;,
\end{equation}
where $\hat H_0$ and $\hat V$ are, respectively, the kinetic energy and interaction terms of the Hamiltonian $\hat H$ in Eq.~(\ref{ham}).  Rewriting the interaction as $\hat V = g \sum_{\bf x} (\hat n_{\bf x}^2 - \hat n_{\bf x})/2$  where $\hat n_{\bf x} = \hat n_{\bf x,\uparrow} + \hat n_{\bf x,\downarrow}$, and expressing $\exp\left(-\Delta\beta g \hat{n}^2_{\bf x}/2\right)$ at each of the $N_{L}^{3}$ lattice points ${\bf x}$ and $N_{\tau}$ time slices $\tau_n=n\Delta\beta$ ($n=1,2,\ldots,N_\tau$) as a Gaussian integral over an auxiliary field $\sigma_{\bf x}(\tau_n)$, the propagator becomes 
\begin{equation}
e^{-\beta \hat{H}} = \int D[\sigma ]G_{\sigma }\hat{U}_{\sigma } + O((\Delta \beta)^2) \;.
\end{equation}
Here $D[\sigma]=  \prod_{{\bf x}, n}\left[d\sigma_{\bf x} (\tau_n)  \sqrt{{{\Delta}\beta |g| /{2\pi}}}\right]$ is the integration measure, $G_{\sigma}=e^{-\frac{1}{2}|g|{\Delta}\beta\sum_{{\bf x}, n}\sigma^{2}_{\bf x}(\tau_n)}$, and $\hat{U}_{\sigma}=\prod_{n}e^{-\Delta \beta \hat H_0/2} e^{-\Delta \beta \hat{h}_{\sigma}(\tau_n)} e^{-\Delta \beta \hat H_0/2}$ (a time-ordered product) with $\hat{h}_{\sigma}(\tau_n)=g\sum_{\bf x} \sigma_{\bf x}(\tau_n)\hat{n}_{\bf x} - g\hat{N}/2$ is the propagator of non-interacting fermions in time-dependent fields $\sigma_{\bf x}(\tau)$.  We use a fast Fourier transform~\cite{Magierski2009, Bulgac2008, Wlazlowski2013} to efficiently change basis between coordinate and momentum space in order to implement the potential and the quadratic single-particle dispersion relation, respectively.  We discretize the integral over each of the $\sigma$ fields using a three-point Gaussian quadrature~\cite{Dean1993}.

The thermal expectation value of an observable $\hat{O}$ is
\begin{equation} \label{expect}
\langle \hat{O} \rangle=\frac{\textrm{Tr}(\hat{O}e^{-\beta \hat{H}})}{\textrm{Tr}(e^{-\beta \hat{H}})}=\frac{\int D[\sigma] \langle \hat{O} \rangle _{\sigma}W_{\sigma}\Phi_{\sigma} }{\int D[\sigma]W_{\sigma}\Phi_{\sigma}} \;,
\end{equation}
where $W_{\sigma}=G_{\sigma}|\textrm{Tr}(\hat{U}_{\sigma})|$, $\Phi_{\sigma}=\textrm{Tr}(\hat{U}_{\sigma})/|\textrm{Tr}(\hat{U}_{\sigma})|$ is the Monte Carlo sign, and $\langle \hat{O} \rangle_\sigma=\textrm{Tr}(\hat{O}\hat{U}_\sigma)/\textrm{Tr}(\hat{U}_{\sigma})$ is the expectation of $\hat O$ with respect to a field configuration  $\sigma$. In AFMC, we sample uncorrelated field configurations according to the positive-definite weight $W_\sigma$ and use them to estimate $\langle \hat O\rangle$ and its statistical fluctuation. 

We project onto fixed particle number $N_s$ for each spin $s$ using the discrete Fourier transform
\begin{equation}\label{particle-projection} 
\hat{P}_{N_s}=\frac{e^{-\beta \mu N_s}}{M}\sum_{m=1}^{M}e^{-i\varphi _{m}N_s}e^{(\beta \mu +i\varphi _{m})\hat{N}_s} \;,
\end{equation}
where $\varphi _{m}=\frac{2\pi m}{M}$ and $M=N_L^3$. 
The chemical potential $\mu$ in (\ref{particle-projection}), chosen to give approximately an average $N_s$, ensures the numerical stability of the Fourier sum. The traces in~\eqref{expect} are computed as canonical traces, 
$\textrm{Tr}_{N_\uparrow,N_\downarrow} \hat X =\textrm{Tr}(\hat{P}_{N_\uparrow} \hat{P}_{N_\downarrow} \hat X)$,  
which are sums of grand-canonical traces using   Eq.~(\ref{particle-projection}). These grand-canonical traces can be computed using the matrix $\mathbf{U}_\sigma$ that represents $\hat U_\sigma$ in the  single-particle space, e.g.,
\begin{equation}
\textrm{Tr}_{\textrm{GC}}[e^{(\beta \mu +i\varphi_{m})\hat{N}}\hat{U}_\sigma]=\textrm{det}[\unit +e^{(\beta \mu +i\varphi _{m})}\mathbf{U}_\sigma] \;.
\end{equation}
We use the diagonalization method of Refs.~\cite{Gilbreth2013, Gilbreth2015} to compute more efficiently the Fourier sums in the number projection.  We also use algorithmic improvements we developed for finite-temperature AFMC calculations of dilute fermionic systems~\cite{Gilbreth2019,supp_material} that have enabled our large-lattice simulations.

\ssec{Results}  We performed AFMC simulations for $N=20,40, 80$ and $130$ particles on lattices of size $7^3, 9^3, 11^3$ and $13^3$, respectively, keeping the filling factor low and constant at $\nu\equiv N/N_L^3 \simeq 0.06$. The ratio of the effective range $r_e \approx 0.337 \, \delta x$~\cite{Werner2012} to the Fermi wavelength is then $k_{\rm F} r_e \simeq 0.41$.  
We use multiple $\Delta\beta$ values for each $\beta$ and a quadratic fit to extrapolate the observables to $\Delta\beta = 0$. For each run, we collect typically between 3,000 and 30,000 thermalized and uncorrelated samples~\cite{supp_material}.  

\begin{figure}[h!]
\begin{center}
\includegraphics[scale=1]{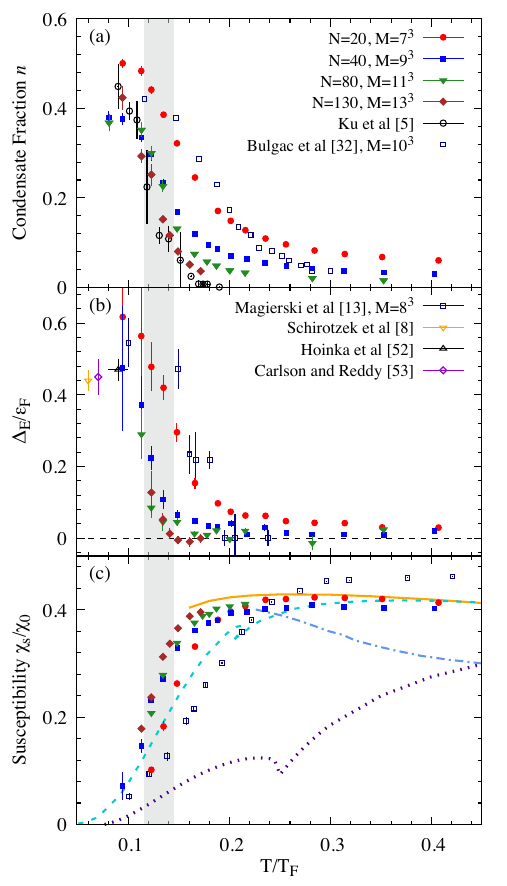}
\end{center}
\caption{(a) AFMC condensate fraction $n$ (solid symbols) compared with experiment~\cite{Ku2012} (open circles) and previous AFMC results~\cite{Bulgac2008} (open squares).  (b) AFMC energy-staggering pairing gap compared with previous AFMC results~\cite{Magierski2009} (open squares) and the low-temperature experiments of Ref.~\cite{Hoinka2017} (open up triangle) and Ref.~\cite{Schirotzek2008} (open  down triangle).  We also compare with the $T=0$ quantum Monte Carlo result of Ref.~\cite{Carlson2008}(open diamond).  (c)  AFMC spin susceptibility compared with the Luttinger-Ward theory~\cite{Enss2012} (solid line), the previous AFMC results of Ref.~\cite{Wlazlowski2013} (open squares), the $t$-matrix results of Ref.~\cite{Palestini2012} (dotted line), the extended $T$-matrix result of Refs.~\cite{Tajima2014,Tajima2016} (dashed line), and the self-consistent NSR results of Ref.~\cite{Pantel2014} (dashed-dotted line). The vertical band is our estimate for $T_c \simeq 0.130(15)\,T_F$.}
\label{fig:ODLRO}
\end{figure}

(i) Condensate fraction: The existence of off-diagonal long-range order in the two-body density matrix $\langle\hat{\psi}^{\dagger}_{\mathbf{k}_1,\uparrow}\hat{\psi}^{\dagger}_{\mathbf{k}_2,\downarrow}\hat{\psi}_{\mathbf{k}_3,\downarrow}\hat{\psi}_{\mathbf{k}_4,\uparrow}\rangle$ is equivalent to this matrix having a large eigenvalue which scales with the system size~\cite{Yang1962}. We calculated the condensate fraction $n$ from the largest eigenvalue $\lambda$, which satisfies $\lambda\leq {N(M-N/2+1)}/(2M) \leq N/2$, using the definition
\begin{equation}
n=\langle \lambda \rangle/[N(M-N/2+1)/(2M)] \;,
\end{equation}
where $M=N_L^3$ is the number of lattice points.
In Fig.~\ref{fig:ODLRO}(a), we show the AFMC condensate fraction for 20, 40, 80, and 130 particles (solid symbols). We compare with the experimental values of Ref.~\cite{Ku2012} (open circles) and the simulations of Ref.~\cite{Bulgac2008} (open squares); for the latter we show the results of the largest lattice reported, $10^3$. 

To obtain the thermodynamic and continuum limits, one must extrapolate to infinite particle number and zero filling factor $\nu \to 0$ or  equivalently $k_F r_e \to 0$~\cite{Burovski2006,Forbes2012,Schonenberg2017}. To determine $T_c$ in the thermodynamic limit at fixed filling factor, we performed a finite-size scaling analysis using the condensate fraction (see Fig.~5 of \cite{supp_material}). We find $T_c \simeq 0.130(15) T_{F}$  for the filling factor of $\nu \simeq 0.06$, shown by the vertical band in Fig.~\ref{fig:ODLRO}.  The Fermi temperature is defined by $T_F=\varepsilon_F/k_B$ where $\varepsilon_F=  (\hbar^2/2m) (3\pi^2 \rho)^{2/3}$ is the Fermi energy and $\rho=\nu/(\delta x)^3$ is the density.  The lower value of $T_c$ at finite filling factor as compared with the experimental value of Ref.~\cite{Ku2012} is consistent with the findings of Ref.~\cite{Gezerlis2008} that the finite effective range of the interaction suppresses the attractive pairing correlations.  The continuum limit requires further studies with large-lattice simulations.  We leave this to future studies.

(ii) Energy-staggering pairing gap:  Using the canonical ensemble, we calculated a model-independent  thermal energy-staggering pairing gap 
\begin{equation}\label{gap}
\Delta_E \! = \! [2E(N_{\uparrow},N_{\downarrow}-1)-E(N_{\uparrow},N_{\downarrow})-E(N_{\uparrow}-1,N_{\downarrow}-1)]/2 \,.
\end{equation}
Here $E(N_{\uparrow},N_{\downarrow})$ is the thermal energy for a system with $N_{\uparrow}$ spin-up particles and $N_{\downarrow}$ spin-down particles. In calculating ({\ref{gap}), we used a particle-number reprojection method~\cite{Alhassid1999,Gilbreth2013}.  This gap does not require a numerical analytic continuation and provides direct information on pairing correlations.  At zero temperature, the energy-staggering pairing gap of the UFG was first studied using quantum Monte Carlo in Ref.~\cite{Carlson2003}.

The pseudogap scenario suggests that pairing correlations appear below a temperature scale $T^{*} > T_{c}$. Such correlations can have various signatures, including a depression in the single-particle density of states, a gap in the single-particle excitation spectrum, and a suppression of the spin susceptibility referred to as ``spin-gap''~\cite{Mueller2017}. If pair formation is energetically favorable, the energy-staggering gap $\Delta_E$ should be nonzero.  However, as shown in Fig.~\ref{fig:ODLRO}(b), $\Delta_E$, which is largely  converged on the $13^3$ lattice near $T_c$,  vanishes above $T_c \simeq 0.130(15)T_{F}$ and does not exhibit a pseudogap signature.

A pairing gap of $\simeq 0.35 \,\textrm{--} \, 0.5 \,\varepsilon_F$ was reported at $T/T_F = 0.15$~\cite{Magierski2009,Magierski2011} (which is  the estimated critical temperature of Ref.~\cite{Bulgac2008}) by fitting the AFMC spectral function to a BCS-like dispersion. Those calculations are shown by the open squares in Fig.~\ref{fig:ODLRO}(b).  
It is unclear whether the gap computed from the spectral function and the gap computed from the energy staggering should agree for the UFG; it would be interesting to perform calculations of these quantities within the same framework.

(iii) Spin susceptibility:  In the presence of pairing correlations, spin-flip excitations require the breaking of pairs, causing a suppression of the spin susceptibility~\cite{Trivedi1995, Kashimura2012, Huscroft2001}.
The spin susceptibility $\chi_s$ is given by
\begin{equation} \label{susceptibility}
\chi_{s}=\frac{\beta}{V}\langle(\hat{N}_{\uparrow}-\hat{N}_{\downarrow})^{2}\rangle \;,
\end{equation}
where the expectation value on the r.h.s. of (\ref{susceptibility}) is calculated for the spin-balanced system $\langle \hat N_\uparrow \rangle  = \langle \hat N_\downarrow \rangle$.
We calculated $\chi_s$ in AFMC using only one particle-number projection onto the total number of particles $N=N_{\uparrow}+N_{\downarrow}$.  In Fig.~\ref{fig:ODLRO}(c) we show our results (solid symbols) for $\chi_s$ in units of the $T=0$ free Fermi gas susceptibility $\chi_{0}=3 \rho/2\varepsilon_F$.  We also compare with the Luttinger-Ward theory of Ref.~\cite{Enss2012} and the $12^3$ lattice AFMC results of Ref.~\cite{Wlazlowski2013}  (open squares), the $t$-matrix results of Ref.~\cite{Palestini2012}, the extended $T$-matrix results of Refs.~\cite{Tajima2014,Tajima2016}, and the Nozi\`eres and Schmitt-Rink (NSR) results of Ref.~\cite{Pantel2014}.

Several calculations found strong suppression of the  spin susceptibility at temperatures above $T_c$ (i.e., at $T/T_F \approx 0.25$ or higher)~\cite{Wlazlowski2013,Palestini2012,Kashimura2012}. This was interpreted as evidence of a pseudogap or a spin gap. In our simulations, we find a suppression of $\chi_s$ only at much lower temperatures close to $T_{c}$ (see also Fig.~9 of Ref.~\cite{supp_material}). For the $13^3, N=130$ system, $\chi_s$ is suppressed for $T \lesssim 0.17\,T_F$.  We note, however, that $\chi_s$ is not fully converged to its thermodynamic limit in our calculations. Since for larger particle numbers, the suppression occurs at lower temperatures, we estimate an upper bound of $T^* \lesssim 0.17 \,T_F$ for the spin-gap temperature at our filling factor of $\nu\simeq 0.06$.  
 We also observe that our large-lattice results agree remarkably well with the theoretical results of Ref.~\cite{Enss2012}.

(iv) Heat capacity: The heat capacity is difficult to compute in quantum Monte Carlo simulations due to large statistical fluctuations. To address this, we use the method of Ref.~\cite{Alhassid2001}, in which the same set of auxiliary fields is used to compute the derivative $C_V= (\partial E/\partial T)_V$, greatly reducing the statistical errors. The heat capacity is shown in Fig.~\ref{fig:HC}  for 20, 40, and 80 particles (solid symbols) along with the experimental results of Ref.~\cite{Ku2012} (open circles). We do not show the 130 particle results since the statistical errors are too large.  We also show in Fig.~\ref{fig:HC} the NSR result of Ref.~\cite{van_Wyk2016}, the diagrammatic $t$-matrix result of Ref.~\cite{Perali2011}, and the Luttinger-Ward results of Refs.~\cite{Zwerger2016,Frank2018}.  Our AFMC results are in overall agreement with the experimental results except for a shift in the peak to a lower temperature.  For the $11^3, N=80$ system the position of the peak at $T \simeq 0.135(10) T_{F}$ is consistent with the value of $T_c \simeq 0.130(15) T_{F}$  for our finite filling factor.  We also note the overall agreement with the Luttinger-Ward results of Refs.~\cite{Zwerger2016,Frank2018}.

Ref.~\cite{van_Wyk2016} described a significant enhancement of the UFG heat capacity at $T \gtrsim T_c$ relative to its value $C_V \approx 1$ in the BEC regime, and attributed this enhancement to metastable preformed cooper pairs present in a pseudogap regime.  Similar enhancement was also observed in the $t$-matrix calculations of Ref.~\cite{Perali2011}. Our calculations confirm such an enhancement. As a function of temperature, we find that it washes out above $T \sim  0.17\, T_F$.

\begin{figure}[tb]
\begin{center}
\includegraphics[scale=0.68]{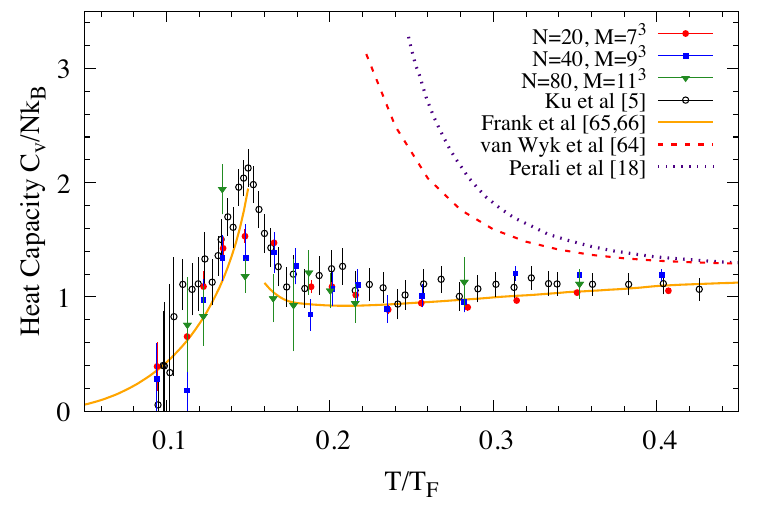}
\end{center}
\caption{The AFMC heat capacity (solid symbols) is compared to experiment~\cite{Ku2012} (open circles), the $t$-matrix result of Ref.~\cite{Perali2011} (dotted line), the NSR result of Ref.~\cite{van_Wyk2016} (dashed line), and the Luttinger-Ward results of Refs.~\cite{Zwerger2016,Frank2018} (solid lines).}
\label{fig:HC}
\end{figure}

\ssec{Model space and spherical cutoff} Signatures of a pseudogap were observed in the AFMC simulations of Refs.~\cite{Magierski2009,Magierski2011,Wlazlowski2013} for temperatures below $\sim 0.25 \,T_F$. Those calculations used a single-particle model space with a spherical cutoff $|{\bf k}| \leq k_{c}=\pi/\delta x$ in momentum. It was shown in Ref.~\cite{Werner2012} that when using such a cutoff, the inverse of the low-momentum scattering amplitude acquires a linear dependence on the center-of-mass momentum ${\hbar\bm{K}}$, and therefore this model does not reproduce the UFG even in the limit $k_F / k_c \rightarrow 0$. 
In Figs.~6-8 of Ref.~\cite{supp_material} we demonstrate this effect in the scattering phase shifts and the two-particle energies. We also checked that our AFMC results change significantly when we introduce a spherical cutoff and become comparable to those of Refs.~\cite{Magierski2009,Magierski2011,Wlazlowski2013}; see Fig.~9 in Ref.~\cite{supp_material}.

\ssec{Conclusion and outlook}
We have presented large-scale AFMC simulations of the homogeneous UFG at a small but finite filling factor of $\nu \simeq 0.06$.  We calculated a model-independent pairing gap $\Delta_E$, the condensate fraction, spin susceptibility, and heat capacity as a function of temperature, and compared these to experiments.

We find that $\Delta_{E}$ vanishes above the critical temperature $T_c$ (which we determine to be $T_{c} \simeq 0.130(15)T_{F}$ for $\nu \simeq 0.06$), and thus does not show pseudogap effects. The spin susceptibility exhibits a moderate spin-gap effect in the range $T_c \lesssim T \lesssim  0.17\, T_F$ for our largest lattice used. 
This result is in contrast to previous AFMC calculations which claim spin-gap effects in a significantly wider range $T_c \lesssim T \lesssim  0.25\, T_F$ at similar filling factors.

Our conclusion holds for $\nu\simeq 0.06$.  It still remains for future work to carry out the continuum extrapolation $\nu \to 0$ or equivalently $k_F r_e \to 0$.  

\ssec{Acknowledgments}  We thank G.F. Bertsch, T. Enss, N. Navon, and F. Werner for useful discussions. We also thank M.J.H. Ku for providing the experimental data of Ref.~\cite{Ku2012}, and  J. Carlson, J.E. Drut,  T. Enss,  B. Frank, P. Magierski, Y. Ohashi, P. Pieri, G. C. Strinati, H. Tajima, G. Wlaz\l{}owski, and W. Zwerger for providing theoretical results shown in Fig.~\ref{fig:ODLRO} and Fig.~\ref{fig:HC} of the main text, and Fig.~4 of the supplemental material. 
This work was supported in part by the U.S. DOE grants Nos.~DE-FG02-91ER40608, DE-SC0019521, and DE-FG02-00ER41132. 
The research presented here used resources of the National Energy Research Scientific Computing Center, which is supported by the Office of Science of the U.S. Department of Energy under Contract No.~DE-AC02-05CH11231.  We also thank the Yale Center for Research Computing for guidance and use of the research computing infrastructure.


%
\vspace{1 cm} 

\setcounter{figure}{0} 

{\bf \large Supplemental material: Pairing correlations across the superfluid phase transition in the unitary Fermi gas}

\subsection{Monte Carlo analysis}

In this section we discuss some technical details of our Monte Carlo analysis.

\subsubsection{Thermalization and decorrelation}
Our auxiliary-field Monte Carlo (AFMC) algorithm works by iterating sequentially through all $N_\tau$ time slices, and for each time slice, updating a fraction of the auxiliary fields and then performing a Metropolis accept/reject. We refer to one iteration through all time slices as a sweep, and to each set of sequential sweeps as a walker.  We generate a number of such independent walkers.  Initially the simulation begins in a region of auxiliary-field configuration space that generically has a small weight in the Hubbard-Stratonovich (HS) path-integral, so some initial sweeps are needed to reach a region of high weight and to ensure that the observables are independent of the initial configuration. This can be seen by plotting the observables, averaged over all walkers, as a function of the sample number in the sequence generated by the Monte Carlo walk. We refer to the initial number of sweeps required to reach this region as the thermalization time $M_0$. In Fig.~\ref{fig:therm} we show an example for the spin susceptibility $\chi_s$. 
\begin{figure}[b]
\includegraphics[scale=1]{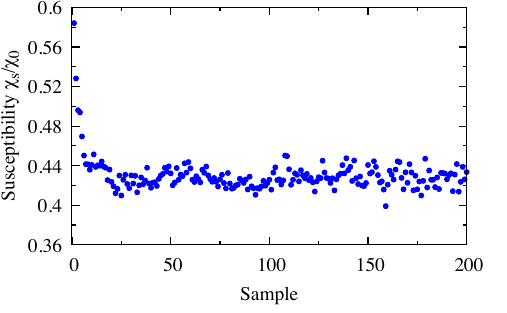}
\caption{Thermalization of the spin susceptibility $\chi_s$ for $N=20$ particles on a $7^3$ lattice at $T/T_F = 0.236$, with $ \varepsilon_F \Delta \beta= 0.028$. Five sweeps were taken between successive samples on the horizontal axis.}
\label{fig:therm}
\end{figure}

Our final observables are calculated using only samples that are obtained after the initial thermalization time. Typically, the logarithm of the partition function $\ln Z$, where $Z = \Tr_{N_\uparrow,N_\downarrow}\hat U_\sigma$ (for spin-up and spin-down particle-number projections) or $\Tr_{N_\uparrow + N_\downarrow} \hat U_\sigma$ (for total particle-number projection), provides the clearest indication of thermalization and is also slower to thermalize than other observables. We therefore use this quantity to determine the thermalization time. For the simulation of Fig.~\ref{fig:therm}, we determined a thermalization time of $M_0=40$  samples or $200$ sweeps. We perform such analysis independently for every AFMC calculation.
\begin{figure}[b]
\begin{center}
\includegraphics[scale=1]{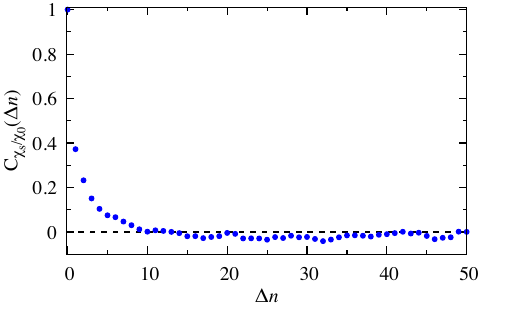}
\end{center}
\caption{Autocorrelation function for the spin susceptibility $\chi_s$ as a function of sample number. Parameters are the same as in Fig.~\ref{fig:therm}. Five sweeps were taken between successive samples on the horizontal axis.}\label{fig:acf}
\end{figure}

 Multiple sweeps are necessary to produce a Monte Carlo sample that is uncorrelated  from the previous one. To determine the number of such sweeps for an observable $\hat X$, known as the decorrelation time, we compute the autocorrelation function $C_{\hat X}(\Delta n)$ of the sequence  $\langle \hat X \rangle_{\sigma^{(n)}}$ for $n >M_0$, where $\sigma^{(n)}$ denotes the auxiliary-field configuration of the $n$-th sample. Typically we compute $C_{\hat X}(\Delta n)$ for each walker and then take its average over all walkers. We then determine the decorrelation time as the number of sweeps $\Delta n$ for which $C_{\hat X}(\Delta n)$ drops below $0.05$.  In Fig.~\ref{fig:acf}, we show the averaged autocorrelation function for the spin susceptibility as a function of sample number, in which 5 sweeps were taken between successive samples. The decorrelation time in the example shown in the figure is $\Delta n = 7$ samples or $35$ sweeps. Typically, $\ln Z$ exhibits a much longer decorrelation time than other observables. While we use $\ln Z$ to determine the thermalization time, we determine the decorrelation time individually for each observable.

The final expectation value $\langle \hat X \rangle$ for a given value of $\Delta \beta$ is computed from $\langle \hat X \rangle = \sum_n \langle \hat X \rangle_{\sigma^{(n)}} \Phi_{\sigma^{(n)}} / \sum_n \Phi_{\sigma^{(n)}}$ using a sequence of thermalized and decorrelated samples, and where $\Phi_{\sigma}$ is the Monte Carlo sign function. We also compute a jackknife estimate of the statistical error of the observable. 

\subsubsection{Extrapolation in $\Delta \beta$}
\begin{figure}[b]
\begin{center}
\includegraphics[scale=1]{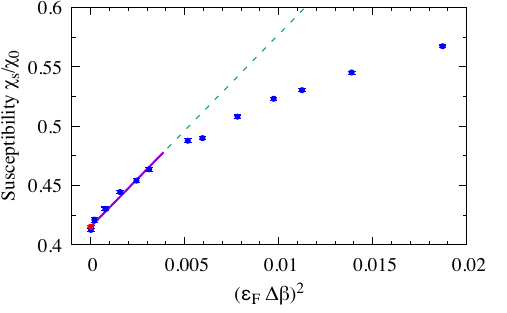}
\end{center}
\caption{Extrapolation in $\Delta \beta$ for the spin susceptibility $\chi_s$ (plotted in units of $\chi_0$). Parameters are the same as in Fig.~\ref{fig:therm}.  The solid circles are the AFMC results, and the solid line is a linear fit in $(\varepsilon_F \Delta \beta)^2$ performed for $(\varepsilon_F \Delta \beta)^2 \leq 0.004$ to obtain the $\Delta \beta = 0$ limit. The extrapolated value is shown by the red circle at $\Delta\beta=0$.}
\label{fig:dbeta}
\end{figure}
The AFMC method is based on a discretized version of the HS transformation, in which the imaginary time $\beta$ is divided into $N_\tau$  discrete time slices $\Delta \beta$. To obtain the final values of the observables it is then necessary to extrapolate their values at finite $\Delta \beta$ to $\Delta\beta=0$.  The error of the Trotter product used in our method scales as $O((\Delta \beta)^2)$ (see Eq.~(4) in the main text). We therefore perform multiple $\Delta \beta$ runs and extrapolate in $(\Delta \beta)^2$ to obtain the $\Delta \beta \rightarrow 0$ limit.  An example for the spin susceptibility $\chi_s$ is shown in Fig.~\ref{fig:dbeta} where $\chi_s/\chi_0$ is plotted as a function of the dimensionless parameter $(\varepsilon_F \Delta\beta)^2$. Typically the quadratic behavior is obtained for $(\varepsilon_F \Delta \beta)^2 \lesssim 0.003$.

\subsection{Algorithmic improvements}

The canonical projection described in Eqs.~(6) and (7) of the main text is computationally time-consuming, even when using the method of Ref.~\cite{SGilbreth2013}, for which the number of floating-point operations required scales as $O(M^3)$, where $M = N_L^3$ or $2 N_L^3$  is the number of states in the single-particle model space. To make the calculations more efficient, we developed several algorithmic  improvements that reduce significantly the effective number of single-particle states for dilute fermionic systems. These techniques, which enabled our large lattice calculations, are discussed in Ref.~\cite{SGilbreth2019}.  As in other finite-temperature AFMC calculations, we store and accumulate the matrix  $\mathbf{U}_\sigma$ as a numerically stabilized matrix decomposition. We decompose $\mathbf{U}_\sigma = Q D R$, where $Q$ is unitary, $R$ is unit upper triangular, and $D$ is diagonal with positive entries. Ref.~\cite{SGilbreth2013} described a method to compute projections using Eq.~(6) of the main text that scales as  $O(N_s^3)$ using this decomposition. For the present work, we extended this method by using the information in the $D$ factor to omit single-particle states which are effectively unoccupied, thereby reducing $N_s$ from the order of thousands to order of hundreds. The error introduced from omitting these states is controlled by an adjustable small parameter, which we tune to ensure our observables are unchanged to $\sim 6$ digits. This reduction is done independently for each field configuration before performing the number projection, which is then done in the reduced model space. Since the number of states in the reduced space scales with the particle number (in a temperature-dependent way) rather than directly the lattice size, the computational time and scaling of the method improves dramatically, making our calculations practical. We implemented similar ideas for optimizing the calculation of certain observables. These methods are described in detail in Ref.~\cite{SGilbreth2019}.

\subsection{Thermal energy}

In addition to the observables described in the main text, we also calculated the thermal energy for the filling factor $\nu\simeq 0.06$.  In Fig.~\ref{fig:Energy} we show (solid symbols), for different particle numbers $N$,  the thermal energy of the UFG in units of the free Fermi gas energy $E_{\rm FG}=\frac{3}{5}N\varepsilon_{\rm F}$ as a function of $T/T_F$. Here $\varepsilon_F=  (\hbar^2/2m) (3\pi^2 \rho)^{2/3}$ and $T_F=\varepsilon_F/k_B$ are the Fermi energy and Fermi temperature of a free Fermi gas with density $\rho=\nu/ (\delta x)^3$. We compare to the experimental results of Ref.~\cite{SKu2012} (open circles), the $T=0$ result of Ref.~\cite{SCarlson2011} for a similar value of the filling factor (open triangle), and the AFMC results of Ref.~\cite{SDrut2012} (open squares). 

\begin{figure}[b]
\begin{center}
\includegraphics[scale=0.65]{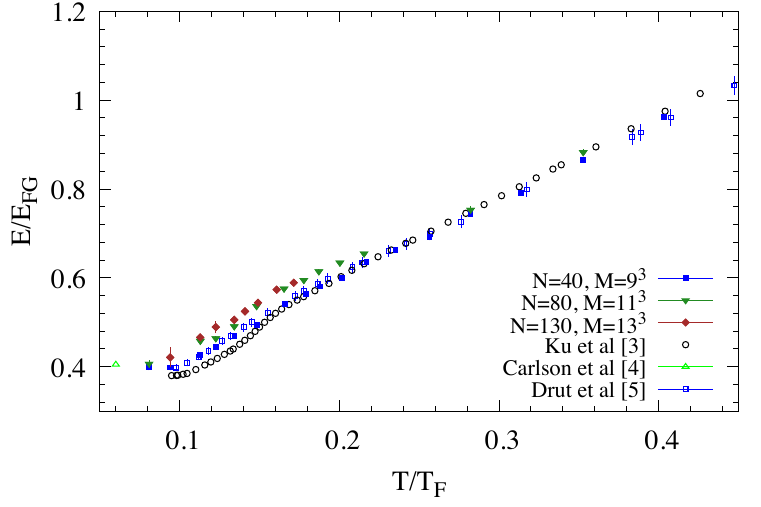}
\end{center}
\caption{AFMC thermal energy (solid symbols) compared to the experiment of Ref.~\cite{SKu2012} (open circles), the ground state result with finite effective range $k_F r_e \simeq 0.38$ of Ref.~\cite{SCarlson2011} (open triangle), and the AFMC results of Ref.~\cite{SDrut2012} (open squares).}
\label{fig:Energy}
\end{figure}

Our results agree with experiment at high temperatures, and are somewhat above the experimental results in the vicinity and below the experimental critical temperature $T_c \simeq 0.17\,T_F$. This is likely due to finite-range effects: at low temperature our result for the thermal energy is comparable with the value of the Bertsch parameter $\xi = {E(T\!=\!0)}/E_{\rm FG}$ calculated in Ref.~\cite{SCarlson2011} to be $\xi \approx 0.405 (2)$ at a similar effective range of $k_{\rm F}r_{e}\simeq 0.38$.  In Ref.~\cite{SCarlson2011} the  Bertsch parameter at zero effective range is estimated to be $\xi = 0.372(5)$, in agreement with experiment~\cite{SKu2012}. See also Ref.~\cite{SEndres2013} where an improved lattice action technique was implemented to obtain $\xi=0.366^{+0.016}_{-0.011}$. 
The AFMC results of Ref.~\cite{SDrut2012}, which also included the complete first Brillouin zone, are slightly lower than our results, possibly due to their lower value of  $k_{\rm F}r_{e}\simeq 0.3$.

\subsection{Finite-size scaling}
To obtain a quantitative estimate of the superfluid critical temperature, finite-size scaling should be performed~\cite{SBinder1981}.  We carried out such finite-size scaling using the condensate fraction data.  The phase transition to superfluidity in the UFG is in the $U(1)$ universality class with the critical exponent $\nu_c\simeq 0.67$~\cite{SGuida1998,SCampostrini2006,SPelissetto2002} associated with the diverging correlation length at the critical point.  Finite-size effects can be captured for the condensate fraction $n(L,T)$ using the scaling ansatz $L^{1+\eta}n(L,T)=f(x)$~\cite{SBurovski2006-2, SGoulko2010}. Here $\eta\simeq 0.038$ for the $U(1)$ class, $x=(L/\xi)^{1/\nu_c}$ with $\xi$ the correlation length, and $f(x)$ is a universal function~\cite{SGuida1998, SCampostrini2006, SPelissetto2002}.  

In the absence of scaling corrections, the condensate fraction data as a function of  $T/T_{F}$ for different particle number $N$ scaled by $L^{1+\eta}$ should cross at the thermodynamic phase transition temperature $T_{c}$.  In Fig.~\ref{fig:FSS} we show the crossing data for the condensate fraction for the simulated filling factor $\nu = 0.06$.  
 For the finite filling factor used in our study, we estimate a critical temperature of $T_{c}=0.130(15) \; T_{F}$ (gray band in Fig.~\ref{fig:FSS}).  The continuum limit $\nu \rightarrow 0$ was addressed in Ref.~\cite{SBurovski2006} where the critical temperature was found to be $T_{c}=0.152(7) \; T_{F}$, and in Ref.~\cite{SGoulko2010} where it was found that $T_{c}=0.173(6) \; T_{F}$.
  \begin{figure}[h]
\begin{center}
\includegraphics[scale=1]{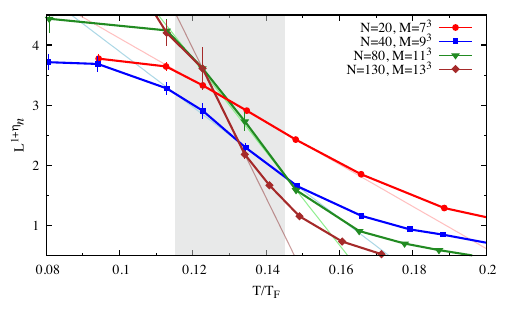}
\end{center}
\caption{Finite-size scaling data for condensate fraction with filling factor of $\nu=0.06$.  The length $L$ is the linear dimension of the lattice $L=7, 9, 11, 13$ for particle numbers $N=20, 40, 80, 130$, respectively.  The critical exponent for a phase transition of the $U(1)$ universality class is taken to be $\eta\simeq0.038$.  We find $T_{c}=0.130(15) \; T_{F}$.}\label{fig:FSS}
\end{figure}

\subsection{Two-particle scattering on the lattice}

To verify the two-particle scattering properties of our lattice model, we solved the scattering problem numerically on the lattice for two particles with varying total center-of-mass (CM) momentum $\hbar\bm{K}$. For each scattering eigenstate, we determined the relative momentum $\hbar k$ from the energy $E$ of the eigenstate using $E=\hbar^2 k^2/2\mu + \hbar^2 K^2 / 2 M$ where $\mu=m/2$ is the reduced mass and $M = 2m$.  We projected onto the $s$-wave component of the wavefunction and determined the $l=0$ phase shift $\delta$ from a fit of the wavefunction to its form $A\left[\cos(\delta) j_0(k r) - \sin(\delta) y_0(k r)\right]$ (where $j_0$ and $y_0$ are, respectively, spherical Bessel functions of the first and second kind) for $r$ larger than the range of the interaction.

At low relative momenta $\hbar k$, the $s$-wave phase shift can be expanded 
\begin{equation}
  k \cot \delta = - a^{-1} + \frac{1}{2} r_{e} k^2 - P r_{e}^3 k^4 + \ldots \,,
  \label{effrange}
\end{equation}
where $a$ is the scattering length, $r_{e}$ is the effective range, and $P$ is the shape parameter.  Using a value for the coupling constant $V_0$  determined by Eq.~(2) in the main text in the unitary limit $a^{-1}=0$, we find that our calculated phase shift fits well the expansion (\ref{effrange}) including terms up to $k^2$  (i.e., for $P=0$). Although the radius of convergence of this expansion is unknown,  we find that the fit parameters $a^{-1}$ and $r_{e}$ are not very sensitive to the range of $k$ used for the fit (as long as the range of $k$ is small enough to not require a shape parameter for a good fit).

In Fig.~\ref{fig:scattering}(a) we show $k \cot \delta$ vs.~$k$ (in units of $1/L$) for the example of a $9^3$ lattice for CM wavevector $K=0$ and  $\bm{K}=2 \pi (1,1,1)/L$. For $K=0$, a fit to (\ref{effrange}) with $P=0$ gives $a^{-1} \approx 0.05 \, L^{-1} \approx 0.006 (\delta x)^{-1}$ and $r_{e} \approx 0.038 \,L \approx 0.34 \, \delta x$ (solid line).  The latter is close to its value $r_{e} = 0.337\ldots \, \delta x$ in the limit of large lattices~\cite{SWerner2012}. The $\bm{K}=2\pi(1,1,1)/L$ points also fit the effective range expansion well for $k \lesssim 15\, L^{-1}$, though deviations due to lattice artifacts are visible for larger values of $k$.  Fig.~\ref{fig:scattering}(b) shows similar results but for a much larger lattice, $25^3$, to illustrate convergence to the continuum limit; here the $k$ dependence is reduced, as expected, and  $r_e \approx 0.015\, L$.

\begin{figure}[t]
\begin{center}
\includegraphics[scale=1]{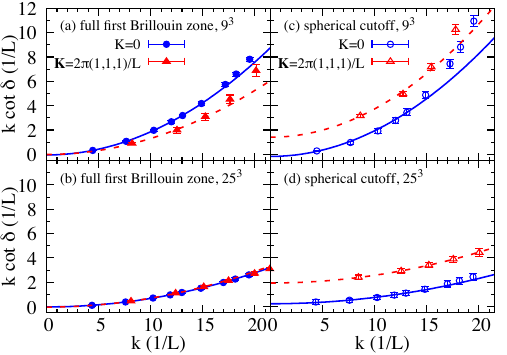}
\end{center}
\caption{$k \cot\delta$ vs.~$k$ for the two-particle scattering on the lattice (symbols).  The phase shift $\delta$ is determined by exact diagonalization for two particles, and the error bars reflect uncertainties in the phase shifts due to lattice artifacts. The lines describe fits of the effective-range expansion (\ref{effrange}) at low momenta. (a) On a $9^3$ lattice, including the full first Brillouin zone of the single-particle momentum. The solid blue line corresponds to CM wavevector $K=0$, and the dashed red line describes $\bm{K}=2 \pi (1,1,1)/L$. (b) On a $25^3$ lattice, also using the full first Brillouin zone. (c) On a $9^3$ lattice, using a spherical cutoff in the single-particle momenta, $|\mathbf{k}| \leq k_c=\pi/\delta x$. (d) On a $25^3$ lattice, also using a spherical cutoff. 
}
\label{fig:scattering}
\end{figure}

\begin{figure}[t]
\begin{center}
\includegraphics[scale=.89]{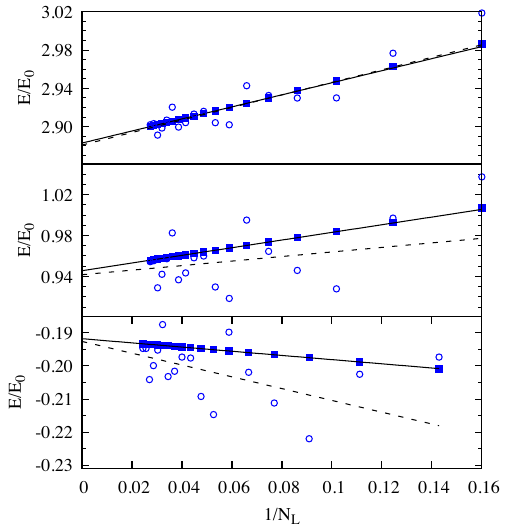}
\end{center}
\caption{Three lowest levels of the two-particle system for zero center-of-mass momentum $\hbar K = 0$ as a function of the inverse number of lattice points $N_L$ in the $x$-direction.  Solid squares show the no-cutoff results (i.e., using the complete cubic first Brillouin zone),  and open circles show the spherical cutoff results. The solid lines are linear fits vs.~$1/N_L$ to the no-cutoff energies and the dashed lines are similar fits to the spherical cutoff energies. In the limit of large lattices the energies calculated with and without a spherical cutoff agree.}\label{fig:A2_K0}
\end{figure}
\begin{figure}[h!]
\begin{center}
\includegraphics[scale=.89]{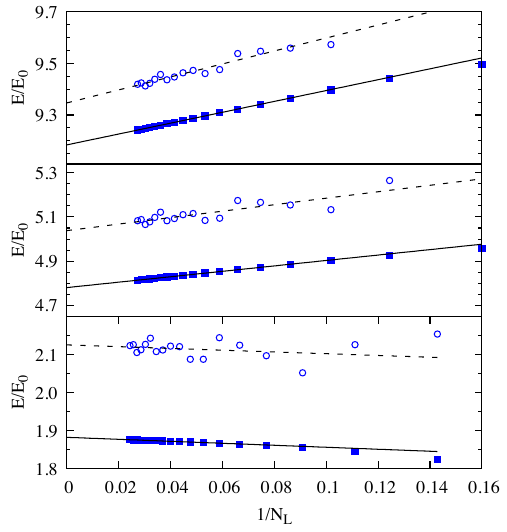}
\end{center}
\caption{As in Fig.~\ref{fig:A2_K0} but for the three lowest levels with center-of-mass wavevector $\bm{K} = 2\pi(1,1,1)/L$. 
 In the limit of large lattices (for which $k_F/k_c \rightarrow 0$), the spherical cutoff energies disagree with the no cutoff energies.}\label{fig:A2_K1}
\end{figure}

\subsection{Spherical cutoff in momentum space}

To determine how a spherical cutoff $|\mathbf{k}| \leq k_c=\pi/\delta x$ in the single-particle momentum space affects the results, we solved the two-particle scattering problem on the lattice as in the previous section but using such a cutoff. 

Our numerical scattering calculations verify that our coupling constant used with the spherical cutoff is correct. We show this in Fig.~\ref{fig:scattering}(c) for the $9^3$ lattice, with the coupling constant $V_0$ determined using Eq.~(2) in the main text in the unitary limit and with ${\cal B}$ being a sphere of radius $k_c$. A fit for $K=0$ to Eq.~(\ref{effrange}) including terms up through $k^2$ yields $a^{-1} \approx 0$.  A better fit can be obtained by including the shape parameter which we omit for simplicity.

In Ref.~\cite{SWerner2012} it was argued that, when using this spherical cutoff, an additional center-of-mass-momentum dependence appears in the two-particle scattering amplitude. This yields an effective range expansion of the form
\begin{equation}\label{effrange_CM}
  k \cot \delta = -a^{-1} + \frac{K}{2 \pi} + \frac{1}{2} r_e k^2 - P r_{e}^3 k^4 + \ldots \;,
\end{equation}
which has an additional term $K/2 \pi$ that persists even in the continuum limit $r_e \rightarrow 0$. Our calculations confirm this effect: Fig.~\ref{fig:scattering}{(c)} shows that on the $9^3$ lattice with a spherical cutoff in the single-particle momentum, center-of-mass wavevector $\bm{K}= 2 \pi (1,1,1)/L$, and the coupling tuned to the unitary limit, the quantity $k \cot \delta$ approaches $ \approx \sqrt{3} L^{-1} = K / 2 \pi$ as $k \rightarrow 0$. This can be seen by inspecting the vertical intercept of the fit (which is close to but not exactly $\sqrt{3}$ due to lattice effects). We demonstrate a similar behavior for a $25^3$ lattice in Fig.~\ref{fig:scattering}(d), showing that the CM momentum $\hbar K$ dependence persists for larger lattices, despite the system otherwise approaching the continuum limit.

This effect also shifts the two-particle energies for nonzero CM momentum. Fig.~\ref{fig:A2_K0} shows the three lowest energies of the two-particle system, in units of $E_{0}=(2\pi\hbar)^2/2mL^2$, at center-of-mass momentum $\hbar K = 0$  as a function of the inverse linear size of the lattice $1/N_L$. Fig.~\ref{fig:A2_K1} shows a similar plot but for $\bm{K}=2\pi(1,1,1)/L$. In these calculations we used the method of Ref.~\cite{SPricoupenko2007} to achieve large lattice sizes. The spherical-cutoff results are noisier due to lattice artifacts related to using a sphere as a cutoff within a cubic momentum lattice. For $K = 0$, the two-particle spherical cutoff energies agree with the full first Brillouin zone energies in the limit of large lattices $1/N_L \to 0$.  We also note agreement with the work of Ref.~\cite{SBeane2004}.  For $\bm{K} = 2\pi(1,1,1)/L$, on the other hand, the spherical cutoff energies disagree with those using the full first Brillouin zone in the limit of large lattices. This limit satisfies $k_F / k_c \rightarrow 0$, demonstrating again that calculations carried out with a spherical cutoff do not reproduce the UFG.

\begin{figure}[b]
\begin{center}
\includegraphics[scale=.95]{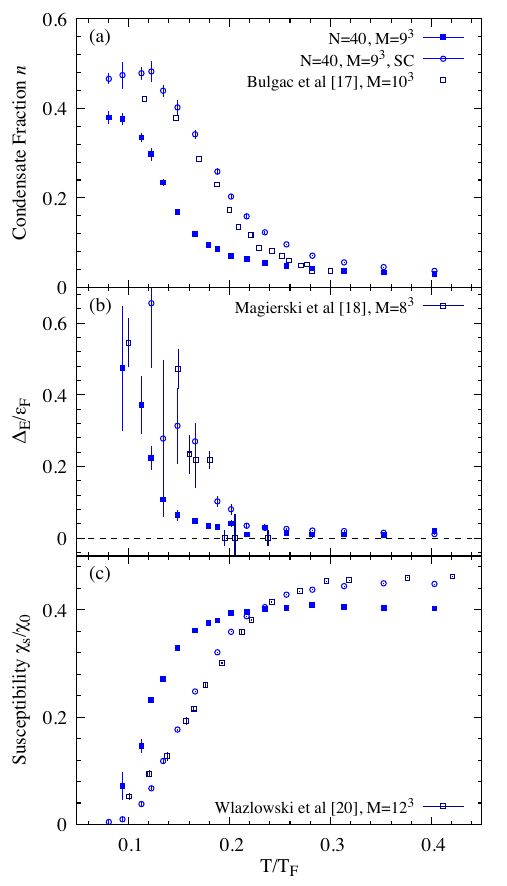}
\end{center}
\caption{AFMC results for (a) condensate fraction, (b) energy-staggering pairing gap, and (c) spin susceptibility vs. $T/T_F$.  Our spherical cutoff results (open circles) for the $9^3$ lattice are compared with the spherical cutoff results of Refs.~\cite{SBulgac2008,SMagierski2009,SMagierski2011,SWlazlowski2013} (open squares) and with our no cutoff results (solid squares).}\label{fig:ODLRO_SC}
\end{figure}

Our AFMC results for the condensate fraction, pairing gap, and spin susceptibility using a $9^3$ lattice for $N=40$ particles, without and with a spherical cutoff in the single-particle momentum space, are shown in Fig.~\ref{fig:ODLRO_SC} (as solid and open circles, respectively).  Our spherical cutoff results (although performed on an odd  lattice)  are comparable to the results of Refs.~\cite{SBulgac2008,SMagierski2009,SMagierski2011,SWlazlowski2013} (shown by open squares), but differ significantly from our AFMC results that use the complete first Brillouin zone in momentum space.  Because the spherical cutoff affects the two-body physics even in the infinite-cutoff limit, this effect cannot be attributed solely to the finite filling factor of our simulation.
For nonzero CM momentum, the energies of the two-particle system with the spherical cutoff are generally higher than without. One therefore expects the system to condense at higher temperatures, which is seen in Fig.~\ref{fig:ODLRO_SC} and is an artifact of the spherical cutoff.

\end{document}